\documentclass{article}
\usepackage{spconf,amsmath,graphicx,url,subfig, cite,booktabs}
\usepackage{graphicx}
\usepackage{amsfonts,amssymb}
\usepackage{xcolor}

\title{Two-stage Neural Network for ICASSP 2023 Speech Signal Improvement Challenge}

\name{
\begin{tabular}{c}
\it Mingshuai Liu$^1$, Shubo Lv$^1$, Zihan Zhang$^1$, Runduo Han$^1$, Xiang Hao$^1$, Xianjun Xia$^2$, \\
Li Chen$^2$, Yijian Xiao$^2$, Lei Xie$^{1*}$
\end{tabular}
}
\address{
  $^1$Audio, Speech and Language Processing Group (ASLP@NPU), School of Software, \\ Northwestern Polytechnical University, Xi'an, China\\
  $^2$ByteDance, China
  }

\begin{document}
\ninept
\maketitle
\begin{abstract}
\vspace{-0.7em}
In ICASSP 2023 speech signal improvement challenge, we developed a dual-stage neural model which improves speech signal quality induced by different distortions in a stage-wise divide-and-conquer fashion. Specifically, in the first stage, the speech improvement network focuses on recovering the missing components of the spectrum, while in the second stage, our model aims to further suppress noise, reverberation, and artifacts introduced by the first-stage model. Achieving 0.446 in the final score and 0.517 in the P.835 score, our system ranks 4th in the non-real-time track.

\end{abstract}

\vspace{-4pt}


\begin{figure*}[!htbp]
	\centering
	\begin{minipage}{0.8\linewidth}
		\centering
		\includegraphics[width=1.0\linewidth]{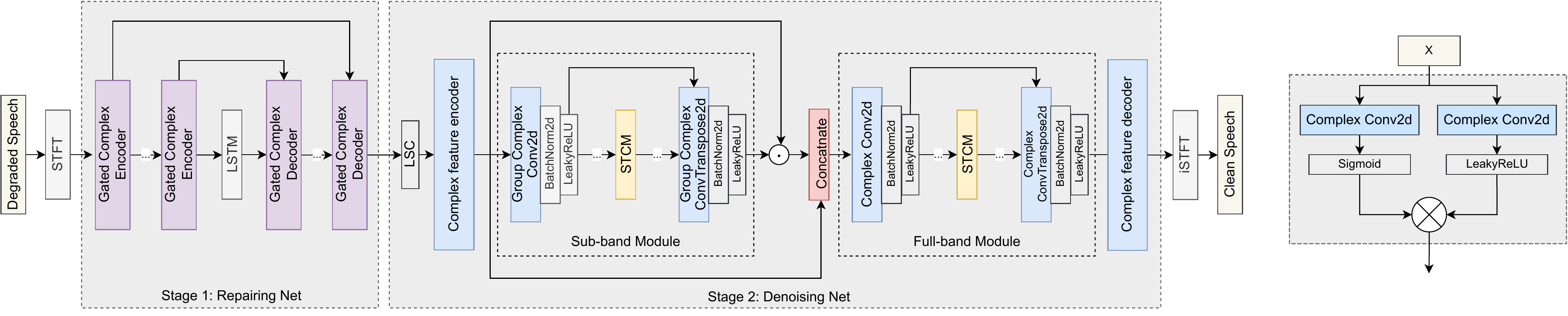}
		\label{f2_1}
	\end{minipage}
    \vspace{-1.5em}
    \caption{Architecture of the proposed two-stage model (left) and gated convolution module (right).}
    \label{f2}
    \vspace{-2em}
\end{figure*}

\vspace{-1em}
\section{Introduction}
\label{sec:intro}
\vspace{-0.8em}


During audio communication, speech signals may be degraded by multiple distortions, including coloration, discontinuity, loudness, noisiness, and reverberation. Existing methods have an impressive performance in noise suppression, but there is still an open problem on how to repair several distortions simultaneously. To stimulate research in this direction, ICASSP 2023 has specifically held the first Speech Signal Improvement Challenge\footnote{https://www.microsoft.com/en-us/research/academic-program/speech-signal-improvement-challenge-icassp-2023/ \ \ \ * Corresponding author.}, as a flagship event of Signal Processing Grand Challenge.



Inspired by the idea of decoupling difficult tasks into multiple easier sub-tasks~\cite{ju2022tea}, we propose a neural speech signal improvement approach with a training procedure that involves two stages.
In the first stage, we adopt DCCRN~\cite{hu2020dccrn} as a repair tool and substitute its oracle convolution structure with a more powerful gated convolution with the aim to mainly repair the missing components in the spectrum. Thus with simulated paired data of perfect and impaired speech, GatedDCCRN learns to improve quality problems caused by coloration, discontinuity, and loudness. For the loss function, besides SI-SNR loss and power-law compressed loss~\cite{wilson2018exploring}, we also integrate adversarial loss further to improve speech naturalness. In the second stage, a variant of S-DCCRN~\cite{lv2022sdccrn} is cascaded to the first stage GatedDCCRN model to particularly remove noise, and reverberation, and suppress possible artifacts introduced by the previous stage. Specifically, S-DCCRN is a powerful denoising model working on super wide-band and full-band signals, consisting of two small-footprint DCCRNs -- one operates on sub-band signal and one works on full-band signal, benefiting from both local and global frequency information. With this denoising model, we further update its bottleneck layers from LSTM to STCM~\cite{li2021two} for better temporal modeling.
The proposed system has achieved 0.446 in the final evaluation score and 0.517 in P.835 score, leading our submission to the 4th place in the non-real-time track.

\section{Approach}
\label{sec:format}
\vspace{-1.1em}

As illustrated in Fig.1, the training procedure of our system consists of two stages. The details of the network architecture, training data, and loss function used in these two stages are introduced below.

\vspace{-1em}
\subsection{Stage 1: Repairing Net}
\vspace{-0.5em}
Considering its impressive ability in signal mapping, we employ DCCRN~\cite{hu2020dccrn} as the backbone of our first-stage speech repair network. DCCRN is a U-Net structured complex network working on complex-valued spectrum, where the encoder and decoder are both composed of layered convolution, and LSTM is served as the bottleneck for temporal modeling. Inspired by the superior performance of gated convolution in image inpainting~\cite{yu2019free}, we update the complex convolution with gated complex convolution, as shown in Fig.1, resulting in GateDCCRN.
For the model training loss, in addition to SI-SNR ($\mathcal{L}_{\text{SI-SNR}}$) and power-law compressed loss ($\mathcal{L}_{\text{PLC}}$), we particularly integrate adversarial loss ($\mathcal{L}_{\text{Adv}}$) by adding the Multi-Period Discriminator~\cite{kong2020hifi} and Multi-Scale Discriminator~\cite{kong2020hifi} into the model optimization process to further improve the speech naturalness. Thus the final loss function is $\mathcal{L}_{\text{stage1}}=\mathcal{L}_{\text{SI-SNR}}+10\cdot\mathcal{L}_{\text{PLC}}+15\cdot\mathcal{L}_{\text{Adv}}.$



\vspace{-1em}
\subsection{Stage 2: Denoising Net}
\vspace{-0.5em}


In the second stage, the pre-trained GateDCCRN and an S-DCCRN~\cite{lv2022sdccrn} are chained as the denoising structure. Specifically, as shown in Fig.1 stage 2, two lightweight DCCRN sub-modules consequently work on sub-band and full-band signals, with the design to model fine details of different frequency bands with further inter-band smoothing. Different from the oracle S-DCCRN, we substitute LSTM with squeezed temporal convolution module (STCM)~\cite{li2021two} in the bottleneck layer of the two DCCRNs, which aims to further strengthen the temporal modeling ability. With this update, the new model is named S-DCCSN. During training, we add noise and reverberation to the data simulated in the first stage to train the second stage model, which makes the model further achieve the ability to suppress noise, reverberation, and artifacts introduced by GateDCCRN. Note that the parameters of the pre-trained GateDCCRN are frozen in this training stage. We adopt SI-SNR loss ($\mathcal{L}_{\text{SI-SNR}}$), power-law compressed loss ($\mathcal{L}_{\text{PLC}}$), and mean square error loss ($\mathcal{L}_{\text{Mag}}$) to optimize the model parameters, and the final loss becomes $\mathcal{L}_{\text{stage2}}=\mathcal{L}_{\text{SI-SNR}}+\mathcal{L}_{\text{PLC}}+\mathcal{L}_{\text{Mag}}$.

\vspace{-1em}
\section{Experiments}
\vspace{-1em}
\label{sec:pagestyle}

\subsection{Datasets}
\vspace{-0.5em}


The training set is created using the DNS4 dataset. Specifically, the DNS4 dataset includes a total of 750 hours of clean speech and 181 hours of noise. In addition, 50,000 RIR clips are simulated by HYB method
. The RT60 of RIRs ranges from 0.2s to 1.2s. The room size ranges from $5 \times 3 \times 3 m^3$ to $8 \times 5 \times 4 m^3$. Totally, there are 110,248 RIR clips, combined from the simulated RIR set and the provided RIR set by the challenge.
We perform dynamic simulations of speech distortions during the training stage. In the first stage, we generate training data degraded by coloration, discontinuity, and loudness, accounting for 60$\%$, 25$\%$, and 15$\%$ respectively. 
For coloration, we follow the description in \cite{liu2022voicefixer} to design a low pass filter. We convolve it with full-band speech and perform resampling on the filtered result to generate low-band speech. Besides producing low-bandwidth distortions, we also restrict signal amplitudes within a range $[-\eta, +\eta]$~$(\eta\in(0, 1))$. When we produce coloration distortion, the low-bandwidth speech and clipping speech account for 60$\%$ and 40$\%$ respectively. Specifically, full band speech is down-sampled to 4K, 8K, 16K, and 24K with the same probability. For discontinuity, the value of speech samples is set to zero randomly with a window size of 20ms. And the probability of setting the value of the sample point to zero is 10$\%$. For loudness, we multiply the signal amplitudes by a scale within a range [0.1, 0.5].
In the second training stage, noise is further added to the first-stage training data with an SNR range of [0, 20]dB, and reverberation is further added to 50$\%$ of the first-stage training data. Dynamic mixing is still used during training. We denote the first-stage and second-stage training data as S1 and S2 respectively, where S2 includes simulations for all signal distortions.

\vspace{-1em}
\begin{table}[htbp]
\centering
\small
 \caption{DNSMOS for ablation models on the official test set.}
 \vspace{-8pt}
\setlength{\tabcolsep}{0.6mm}
 \label{tab:dnsmos}
  \resizebox{\linewidth}{!}{
\begin{tabular}{@{}lccccccccccccccccc@{}}
\toprule
    Model & & & & Data & & Para.~(M) & & & SIG & & & & & BAK & & & OVRL \\ \midrule
Noisy & & & & - & & - & & & 2.89 & & & & & 3.45 & & & 2.46       \\ 
DCCRN & & & & S1 & & 5.05 & & & 2.93 & & & & & 3.48 & & & 2.53      \\ 
GateDCCRN w/o Disc. & & & & S1 & & 6.70 & & & 2.96 & & & & & 3.49 & & & 2.55      \\
GateDCCRN & & & & S1 & & 6.70 & & & 3.02 & & & & & 3.53 & & & 2.60    \\
GateDCCRN + S-DCCRN & & & & S1$\rightarrow$S2 & & 9.70 & & & 3.16 & & & & & 3.94 & & & 2.87      \\
GateDCCRN + S-DCCSN & & & & S1$\rightarrow$S2 & & 10.00 & & & \textbf{3.21} & & & & & \textbf{4.02} & & & \textbf{2.90}      \\
GateDCCRN & & & & S2 & & 6.70 & & & 3.05 & & & & & 3.61 & & & 2.64      \\
S-DCCSN & & & & S2 & & 3.30 & & & 3.10 & & & & & 3.78 & & & 2.70      \\
\bottomrule
\end{tabular}
}
\vspace{-2.5em}
\end{table}

\begin{table}[htbp]
\centering
\tiny
 \caption{Multi-dimensional subjective results on the official test set.}
 \vspace{-8pt}
\setlength{\tabcolsep}{0.6mm}
 \label{tab:corrcoef}
  \resizebox{\linewidth}{!}{
\begin{tabular}{@{}lcccccccc@{}}
\toprule
    & Score & Overall & Signal & Noise & Coloration & Discontinuity & Loudness & Reverb \\ \midrule
Noisy & 0.411 & 2.360 & 2.927 & 3.302 & 3.029 & 4.061 & 2.992 & 3.852 \\ 
GateDCCRN+S-DCCSN & 0.446 & 2.608 & 2.964 & 4.058 & 3.131 & 3.673 & 3.601 & 4.335 \\
\bottomrule
\end{tabular}
}
\vspace{-2.5em}
\end{table}

\vspace{-1.1em}
\subsection{Experiment Setup}
\vspace{-0.5em}
The window length and frame shift are 20ms and 10ms respectively, resulting in 10ms algorithmic latency and 10ms buffering latency. The STFT length is 1024. 
The number of channels for DCCRN~/~GateDCCRN is $\{$16, 32, 64, 128, 256, 256$\}$, and the convolution kernel size and stride are set to (5,2) and (2,1) respectively. There are two LSTM layers with 256 nodes followed by a 2048 $\times$ 256 fully connected layer between the encoder and decoder.
The number of channels for the sub-band module and the full-band module of S-DCCRN~/~S-DCCSN is $\{$64, 64, 64, 64, 128, 128$\}$, and the convolution kernel size and stride are set to (5,2) and (2,1) respectively. For the CED and CFD of S-DCCRN~/~S-DCCSN, the number of channels is 32 and the depth of DenseBlock is 5. In S-DCCSN, The hidden channels of STCM adopted by the sub-band module and full-band module are 64. While in S-DCCRN, there are two LSTM layers with 256 nodes instead. And a fully connected layer with 512 nodes is adopted after the last LSTM layer. Models are optimized by Adam with the initial learning rate of 0.001, halved if the validation loss of two consecutive epochs no longer decreases. 

\vspace{-1em}
\subsection{Results}
\vspace{-0.6em}
We conduct ablation experiments to validate each proposed module.
In Table 1, we train DCCRN and GateDCCRN with and without (w/o) discriminator using the first-stage training data (S1). Then we froze the parameters of the pre-trained GateDCCRN and cascade this model with S-DCCRN and S-DCCSN respectively. And the cascaded models are trained using the second-stage training data (S2). In addition, we also train a GateDCCRN and an S-DCCSN with the second-stage training data only (S2). 
DNSMOS results in Table 1 show that gated convolution and adversarial training using a discriminator can effectively improve the signal improvement ability of DCCRN. Moreover, S-DCCSN surpasses S-DCCRN in all evaluation metrics with a small increase in model size, which proves the better temporal modeling ability of STCM. 
Finally, the single-stage models (GateDCCRN and S-DCCSN) trained using S2 reflecting all distortions are inferior to the multi-stage model (GateDCCRN+S-DCCSN). 
Table 2 shows the subjective results of our submitted two-stage model (GateDCCRN+S-DCCSN) on the official test set. We can see that speech quality is clearly improved for different types of distortions except for discontinuity. The bad cases may attribute to our imperfect data simulation on discontinuity which desires further investigation.
The number of parameters of the submitted two-stage system is 10M. The RTF is 1.478 tested on Intel(R) Xeon(R) CPU E5-2678 v3 2.4GHz using a single thread. 

\footnotesize
\bibliographystyle{IEEEbib}
\let\oldbibliography\thebibliography 
\renewcommand{\thebibliography}[1]{ 
  \oldbibliography{#1}
  \setlength{\itemsep}{-1pt} 
}
\bibliography{strings,refs}

\end{document}